# Numerical-aperture transfer in holotomography with a deterministic diffusion prior

DONG-MIN RYU,[1,†] JEONG-HYUN ROH,[1,†] WOON-SOO LEE,[1]
HANSOL YOON,[1,3] HYUN-SEOK MIN,[1,4] AND YONGKEUN PARK[2,*]

[1]Tomocube Inc., Daejeon 34127, Republic of Korea
[2]Department of Physics, Korea Advanced Institute of Science and Technology (KAIST), Daejeon 34141, Republic of Korea
[3]Current Affiliation: Electrical and Computer Engineering, Georgia Institute of Technology, Atlanta, Georgia 30332, USA
[4]e-mail: hsmin@tomocube.com
[†]These authors contributed equally.
*yk.park@kaist.ac.kr

Abstract: High-throughput holotomography often relies on long-working-distance, multiwell-compatible optics that reduce illumination numerical aperture (NA) and limit access to high spatial frequencies. Here we present ResShift-ODE, a deterministic diffusion-prior framework that transfers low-NA refractive-index (RI) tomograms to high-NA-equivalent volumes without modifying the acquisition hardware. We formulate low-NA-to-high-NA transfer as a diffusion-prior inverse problem under an explicit NA-limited Fourier-domain forward operator, distinct from suppressing artifacts within an already measured passband. The method extends residual-shifting diffusion to volumetric RI data and reformulates the reverse process as a probability-flow ordinary differential equation, enabling reproducible inference in five denoiser evaluations. On held-out emulated-pair test volumes, inferred volumes matched high-NA references with RI errors of 0.002–0.003 for >99% of voxels, while Fourier analysis confirmed measurement-anchored lateral-band recovery without filling the axial missing cone. 3D ResShift-ODE required five denoiser evaluations per volume, incurring ~5.8× the cost of a 3D U-Net while remaining ~166× faster than a 1000-step 3D denoising diffusion probabilistic model.

## 1. Introduction

Holotomography (HT) reconstructs the three-dimensional (3D) refractive index (RI) distribution of live biological specimens from multiple 2D measurements of diffracted light fields [1–4]. Because RI is an intrinsic optical property that correlates with local dry mass and molecular composition [5,6], HT provides quantitative, label-free 3D contrast without exogenous fluorophores, stains, or genetic reporters. This capability has enabled applications spanning organelle and lipid-droplet imaging [7–9], biomolecular condensates [10–12], organoids [13,14], histopathology [15,16], and microbiology [17,18]. These strengths make HT an attractive label-free complement to fluorescence high-content imaging.

High-throughput HT faces an inherent trade-off between screening-compatible optics and subcellular RI resolution. Research-grade systems can pair confocal-dish formats with short-working-distance objectives to achieve high illumination and detection NAs, yielding broad lateral bandwidth and improved axial sectioning [19]. By contrast, multiwell plates, microfluidic devices, and organoid screens generally require long-working-distance, open-top optics that limit illumination NA, often to <0.5. This reduced synthetic-aperture coverage truncates the accessible 3D Fourier support of the scattering potential [3,19], suppressing high-spatial-frequency RI content, broadening fine structures, and lowering apparent RI contrast.

Computational approaches have partially addressed related limitations in 3D RI tomography. DeepRegularizer uses a 3D U-Net to map low-quality tomograms to total-variation-regularized reconstructions, improving reconstruction speed and apparent image quality [20]. Projection-domain cycle-consistent learning suppresses missing-cone artifacts without paired supervision [21], and SnSNet combines a learned support mask with Gerchberg–Papoulis iteration for limited-angle HT [22]. Physics-based strategies, including partial-volume Rytov reconstruction with wave back-propagation, extend the accessible depth range of RI imaging [23]. Learned-prior reconstruction has also been applied in HT, for example by embedding a deep-learning projector inside the reconstruction as a regularizing prior [24]. These methods are valuable, but they are not designed for NA transfer: recovering high-NA-equivalent 3D RI volumes from low-NA measurements under a known NA-limited Fourier-domain forward operator. Because the low-NA synthetic aperture truncates the scattering-potential spectrum, the missing high-frequency band is not simply blurred but absent from the measurement. The task is therefore band completion rather than deconvolution, requiring a learned prior that estimates unmeasured frequency content while remaining anchored to the measured low-NA support.

Diffusion models provide a natural framework for such prior-guided inverse problems. In microscopy, denoising diffusion probabilistic models (DDPMs) [25] have been adapted into physics-informed variants that incorporate point-spread-function models for 2D image reconstruction [26], and Brownian-bridge diffusion has enabled pixel super-resolution for 2D virtual staining of label-free tissue images [27]. Related score-based approaches have also been explored for optical inverse problems and inverse scattering [28,29]. However, these demonstrations are primarily 2D, rely on incoherent or simplified forward models, or address modality translation rather than native volumetric RI recovery. Diffusion solvers for inverse problems under a known operator are by now established [30–33]. However, low-NA-to-high-NA transfer of real 3D RI volumes has not been formulated as a diffusion-prior inverse problem under an explicit NA-limited Fourier-support operator with measurement-anchored, missing-cone-preserving behavior. That operator-plus-modality formulation, not the diffusion-inverse-problem paradigm itself, is what we claim as new.

Here we present ResShift-ODE, a volumetric diffusion-prior framework with deterministic probability-flow ODE sampling for NA transfer in HT. The method extends residual-shifting diffusion [34] to volumetric RI data by anchoring the reverse trajectory to the low-NA measurement rather than to an unconditional Gaussian endpoint. We further reformulate the stochastic reverse process as a probability-flow ODE, enabling reproducible few-step inference with the same trained denoiser. We generate voxel-registered supervision by applying an analytic low-NA forward operator to experimentally acquired high-NA reconstructions and evaluate the resulting model across yeast, HepG2 cells, and K562 lymphoblasts. On held-out emulated-pair test volumes, ResShift-ODE achieves high-NA-consistent RI recovery with five-step deterministic inference.

## 2. Methods

### 2.1 Physical formulation of numerical-aperture transfer

We formulate NA transfer in HT as conditional recovery of a high-NA RI volume from a low-NA RI observation (Fig. 1). Let $x_H$ denote a high-NA RI tomogram reconstructed under confocal-dish-class optics with illumination (NA = 0.72), and let $x_L$ denote the corresponding low-NA observation under multiwell-compatible optics with illumination (NA = 0.38). In Fourier space, the low-NA support $K_L$ is contained within the high-NA support $K_H$, while the high-NA configuration contains additional spatial-frequency information outside $K_L$. The goal is thus recovery of high-NA-consistent RI structure while preserving the measured low-NA content, not ordinary deblurring.

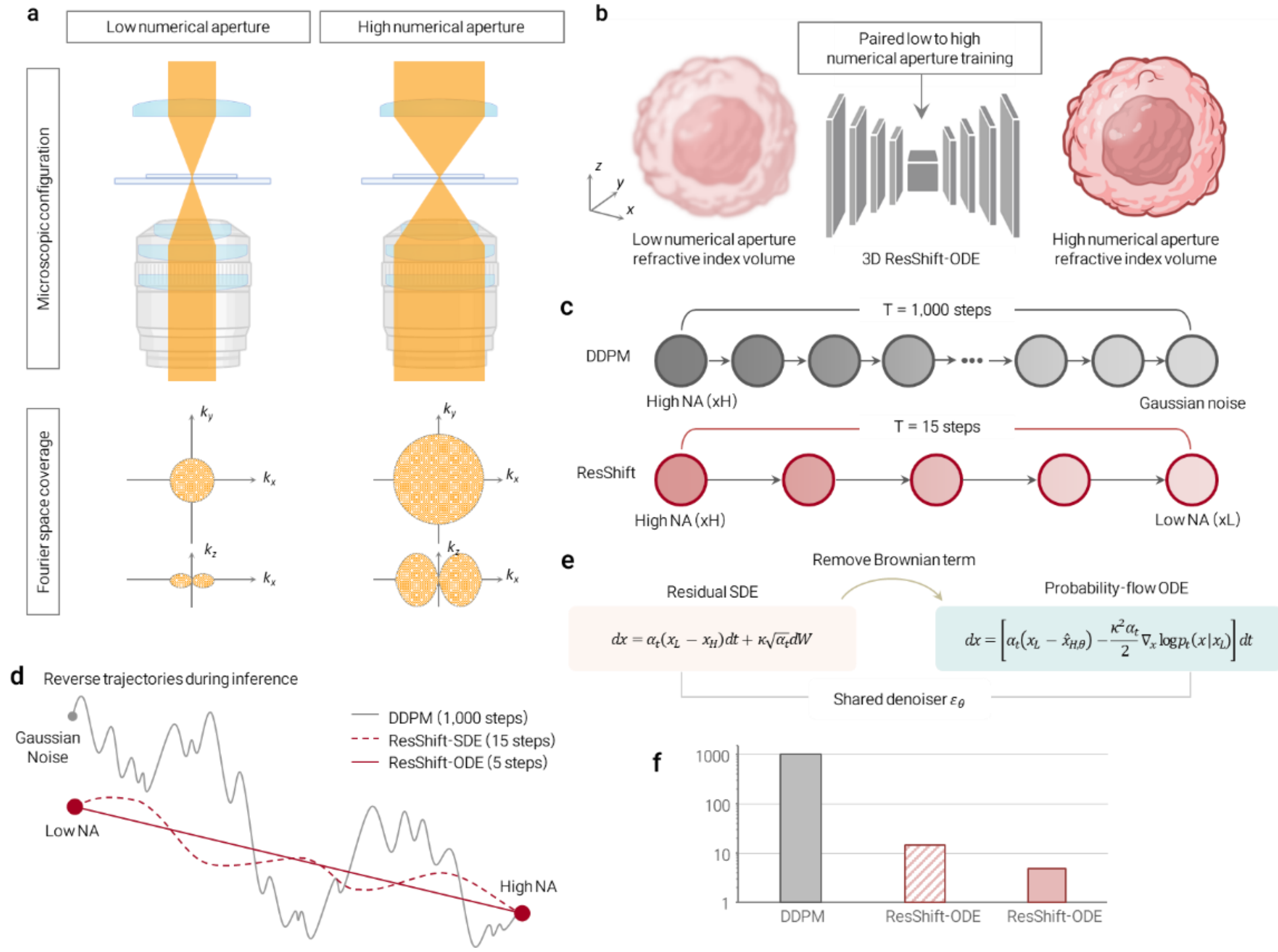

**Fig. 1. Concept and algorithm of ResShift-ODE for NA transfer in HT.** (a) Low-NA illumination restricts Fourier-space coverage, reducing lateral bandwidth and increasing axial elongation, whereas high-NA illumination expands the synthetic aperture. Schematic Fourier supports are shown below each configuration. (b) Training workflow: paired low-NA inputs are generated from measured high-NA RI volumes using the NA-limited forward operator, and the trained model maps low-NA tomograms to high-NA-equivalent outputs. (c) Forward processes of DDPM and ResShift: DDPM diffuses toward Gaussian noise, whereas ResShift shifts toward the paired low-NA measurement. (d) Reverse trajectories for DDPM, stochastic ResShift, and deterministic ResShift-ODE. (e) The probability-flow

ODE removes the Brownian term while reusing the trained denoiser without retraining. (f) DDPM, stochastic ResShift, and ResShift-ODE require 1000, 15, and 5 denoiser evaluations, respectively.

To first order, the low-NA observation is modeled as a coherent Fourier-support restriction of the high-NA RI volume,

$$\boldsymbol{x}_{\mathrm{L}} = \mathcal{F}^{-1}\left[\boldsymbol{M}_{\mathcal{K}_L}\,\mathcal{F}[\boldsymbol{x}_{\mathrm{H}}]\right] + \boldsymbol{n}_{\mathrm{obs}}, \tag{1}$$

where $\mathcal{F}$ is the 3D Fourier transform, $\boldsymbol{M}_{\mathcal{K}_L}$ is the support mask corresponding to the low-NA synthetic aperture, and $n_{obs}$ represents residual reconstruction noise. The mask encodes the frequency region accessible under the low-NA illumination cone and detection pupil, following the Fourier-diffraction geometry of HT [4,19].

The low-NA-to-high-NA inverse mapping is ill posed because the NA-limited forward operator in Eq. (1) removes spatial-frequency components outside $\mathcal{K}_L$, leaving them unconstrained by the low-NA measurement. Many high-NA RI volumes can therefore be consistent with the same low-NA observation. We address this non-uniqueness with a conditional 3D diffusion prior anchored to the measured low-NA input; the recovered high-frequency content is prior-guided completion constrained by the measurement, not directly measured information.

### *2.2 Numerical low-NA emulation and paired-data generation*

Voxel-registered supervision is obtained by applying Eq. (1) to experimentally acquired high-NA RI reconstructions. Each high-NA volume is transformed into Fourier space, multiplied by the low-NA synthetic-aperture support mask, and inverse-transformed; the cropped volume is then passed through the same non-negativity, and total-variation–regularized reconstruction used for the high-NA data, producing an emulated low-NA input on the same reconstruction grid. Because this final regularized reconstruction is nonlinear, the realized emulated volume is not strictly zero outside the low-NA support; the weak object-correlated content beyond that support in Fig. 2(d) is reconstruction-induced rather than measured. Emulating the full measurement-and-reconstruction chain, rather than the optical band limit alone, makes the emulated input a closer surrogate for a physically acquired low-NA reconstruction. This low-from-high strategy follows supervised restoration paradigms in microscopy [35–37], but the degradation here is the Fourier-space support of low-NA HT. The resulting pairs avoid the lateral drift and field-of-view mismatch inherent to two-shot physical acquisitions of the same specimen.

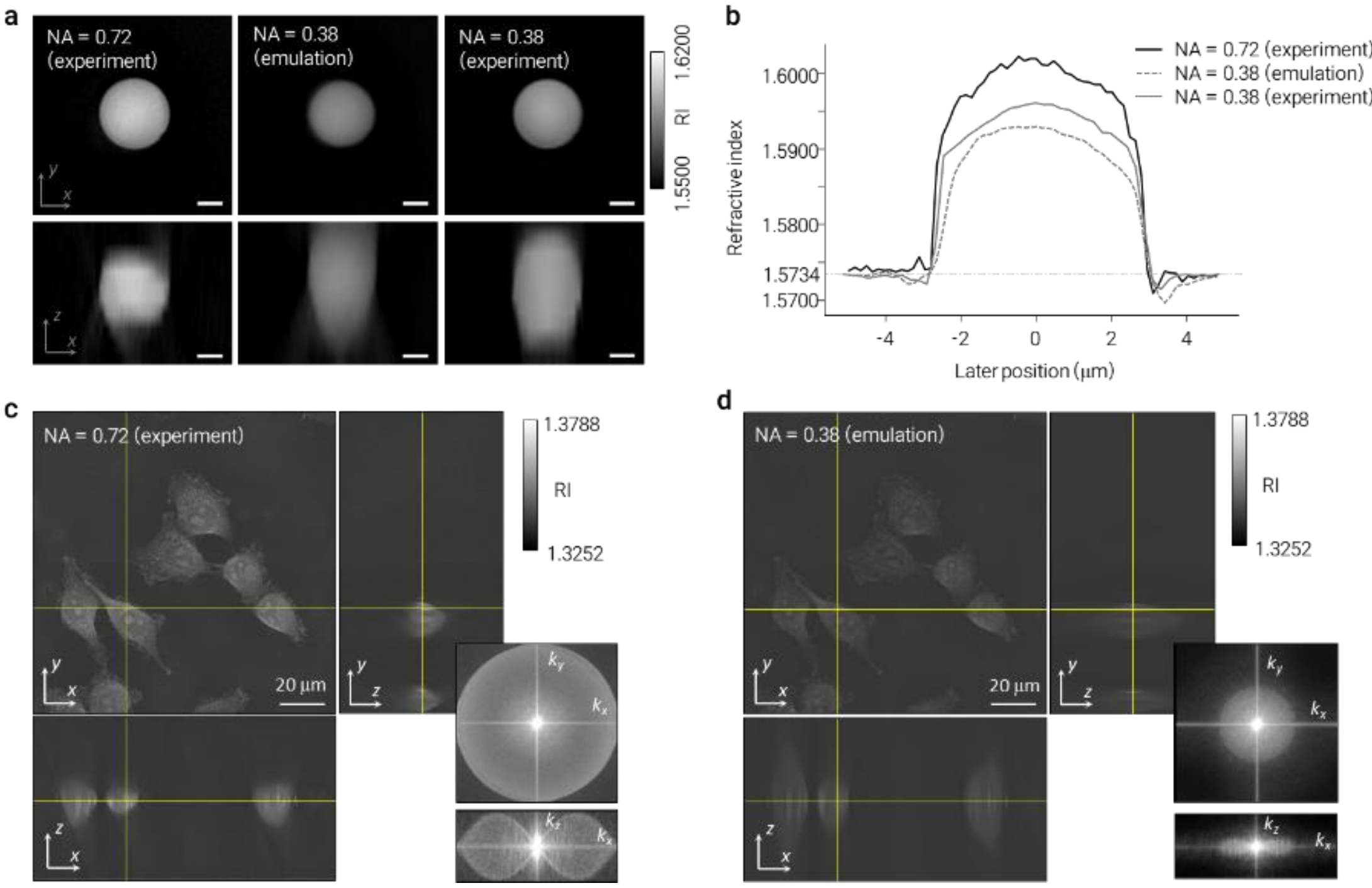

**Fig. 2**. Validation and use of numerical low-NA emulation for paired-data generation. (a) A high-NA bead tomogram acquired at illumination NA = 0.72 was converted to an emulated low-NA tomogram using the NA-limited forward operator for illumination NA = 0.38 and compared with an independently acquired low-NA measurement. (b) Bead-center RI line profiles show close agreement between emulated and measured low-NA profiles, supporting use of the operator for voxel-registered paired-data generation. (c) Representative high-NA HepG2 RI tomogram shown as orthogonal sections with Fourier-domain amplitude projections. (d) Emulated low-NA version of the same tomogram, showing reduced Fourier support, attenuated subcellular contrast, and axial blurring while preserving registration with the high-NA reference.

We validate the emulation operator using a calibrated polystyrene-bead phantom by comparing an emulated low-NA volume generated from a high-NA measurement with an independently acquired low-NA measurement (Fig. 2a,b). The two low-NA profiles show similar lateral broadening, axial elongation, and reduced peak RI contrast, supporting the use of the analytic operator for paired-data generation. The support mask is a best-case (sharp, lossless-in-band) model of the low-NA synthetic aperture and omits pupil apodization, partial coherence, and aberration; it therefore represents an upper bound on operator fidelity rather than a complete physical model. The same operator is then applied to high-NA reconstructions of Saccharomyces cerevisiae yeast, HepG2 adherent cells, and K562 lymphoblasts to construct the multi-cell-type training corpus (Fig. 2c,d). This protocol validates inversion of a known operator. The corresponding Fourier projections show the reduced synthetic-aperture support relative to the high-NA measurement, and each voxel-registered high-NA/low-NA pair serves as a training datum.

### 2.3 ResShift-ODE diffusion prior

ResShift [34] replaces the Gaussian endpoint of a standard DDPM with a residual-shifting Markov chain whose endpoint is the low-NA measurement, anchoring the reverse trajectory to the measured input rather than to an unconditional Gaussian. We extend this process to 3D RI tomograms and use the trained denoiser as a conditional prior for recovering high-NA-consistent structure; the discrete forward process and transition kernel are given in Supplementary Section S1.

To obtain deterministic and reproducible inference, we reformulate the stochastic residual-shifting dynamics as a probability-flow ODE [38–40]. This deterministic sampler is not new machinery: under the reparameterization $s \equiv \eta_t$ it is equivalent to the linear conditional Flow-Matching and rectified-flow path [41,42], and is related to the I²SB Schrödinger-bridge construction [43]. Our contribution is therefore not the sampler but the problem formulation, NA transfer under an explicit NA-limited operator, and the measurement-anchored, cone-preserving behavior it produces. Treating the discrete chain as the Euler–Maruyama discretization of an underlying stochastic differential equation gives the continuous-time forward process, in which $\kappa$ is the scalar noise scale of the residual-shifting kernel (in RI units; its numerical value and the $\eta_t$ schedule are given in Supplementary Section S2)

$$d\boldsymbol{x} = \alpha_t(\boldsymbol{x}_\mathrm{L} - \boldsymbol{x}_\mathrm{H})\,dt + \kappa\sqrt{\alpha_t}\,d\boldsymbol{W}, \qquad \alpha_t = \frac{d\eta_t}{dt}. \tag{2}$$

In the discrete chain, $\alpha_t = \eta_t - \eta_{t-1}$ corresponds to $d\eta_t/dt$ under unit-step ($\Delta t = 1$) Euler–Maruyama discretization, with $\eta_0 = 0$ and $\eta_T = 1$; integrating Eq. (2) from $x_0 = x_H$ then transports the state to $x_T = x_L$. Concretely, the reverse trajectory satisfies,

$$d\boldsymbol{x} = \left[\alpha_t\big(\boldsymbol{x}_\mathrm{L} - \hat{\boldsymbol{x}}_{\mathrm{H},\theta}(\boldsymbol{x}_t, t)\big) - \kappa^2\alpha_t/2\ \nabla_{\boldsymbol{x}}\log p_t(\boldsymbol{x} \mid \boldsymbol{x}_\mathrm{L})\right] dt. \tag{3}$$

The network predicts the clean high-NA volume from the current ODE state and the low-NA condition, with the same conditioning as the trained denoiser $\hat{x}_{H,\theta}(x_t, t, x_L)$ defined in Eq. (S5). Through Tweedie's identity, the score term in Eq. (3) is itself expressed through the denoiser output $\hat{x}_{H,\theta}$, so the bracketed drift reduces to a single function of that output and each Euler step requires exactly one denoiser evaluation (Supplementary Section S1). We integrate the reverse ODE with a first-order Euler scheme at five uniformly spaced steps (number of function evaluations, NFE = 5). The resulting trajectory is deterministic for fixed initialization and conditioning, giving reproducible output volumes. This determinism improves reproducibility, not ground-truth certainty: accuracy in the unmeasured high-frequency band remains bounded by the learned prior and by the validity of the low-NA forward model.

Equivalence note: under the reparameterization $s \equiv \eta_t$, the probability-flow ODE—the deterministic mean trajectory—coincides with the linear conditional Flow-Matching path $\boldsymbol{x}(s) = (1 - s)\,\boldsymbol{x}_\mathrm{H} + s\,\boldsymbol{x}_\mathrm{L}$ [41,42]. The equivalence is to the deterministic ODE, not to the stochastic forward SDE of Eq. (2). We adopt the residual-shifting derivation here for direct continuity with the 2D ResShift lineage [34], and cite the Schrödinger–Bridge framing [43] for completeness without claiming priority.

### 2.4 Network architecture

The denoiser network uses a customized 3D U-Net backbone (a factorized, pseudo-3D design rather than dense Conv3D). The input is the channel-wise concatenation of the current diffusion state and the low-NA condition, and the time step is injected through sinusoidal embeddings into each residual block.

Three architectural choices distinguish our 3D backbone from a vanilla volumetric U-Net: (i) a factorized Conv2D + axial-attention core—2D convolutions per slice, axial self-attention across slices—chosen for tractable 3D memory and compute; (ii) pre-activation residual blocks (GroupNorm → SiLU → conv) for stable deep-stack training; and (iii) relative position bias in the attention modules with rotary position embedding on the channel axis. Full layer schedule, channel widths, patch size, and tiling are given in Supplementary Section S2.

### 2.5 Training data

The model is trained on a deliberately multi-cell-type corpus of paired low-NA/high-NA RI volumes acquired on a HT system (HT-X1, Tomocube Inc., Daejeon, Republic of Korea) spanning three morphological regimes—suspended yeast, adherent HepG2, and organelle-rich K562. Each class contributes 20 paired volumes generated from experimentally acquired high-NA reconstructions by the numerical low-NA operator; dataset splits, acquisition, normalization, augmentation, and training hyperparameters are given in Supplementary Section S3.

## 3. Results

### 3.1 Evaluation scope and test protocol

The quantitative experiments use emulated low-NA inputs with voxel-registered high-NA references. The corpus comprises 60 paired low-NA/high-NA volumes (20 per class), split into 51 training, 6 validation, and 3 test volumes, with one held-out test volume per class (HepG2, K562, and Yeast). All quantitative metrics below should therefore be read as per-volume operator-inversion results on held-out specimens, not as population-level statistics. Reconstruction quality is quantified with PSNR, SSIM [44], and a per-voxel-normalized Laplacian mean-squared error (L-MSE); metric definitions are given in Supplementary Section S4.

The four compared models, 3D U-Net, 3D DDPM, 3D ResShift, and the proposed ResShift-ODE, share the same backbone architecture and training data; the diffusion training loss is shared only among the three diffusion models, as the 3D U-Net is a deterministic regressor. The proposed model further differs from the stochastic 3D ResShift baseline in the residual-shifting schedule (shape and coefficients; Section S2), not only in the reverse sampler, so the comparison reflects the combined effect of the diffusion formulation, schedule, and sampler rather than the sampler alone.

### 3.2 Cross-cell-type high-NA recovery

We first ask whether the model restores high-NA-like RI structure across biologically distinct morphologies. Figures 3a-c compare the low-NA input, inferred high-NA output, and experimentally measured high-NA reference for yeast, HepG2, and K562 lymphoblasts using orthogonal *xy*, *yz*, and *xz* sections. The low-NA inputs show broadened cell envelopes and attenuated subcellular contrast, whereas the high-NA references resolve class-characteristic RI texture—vacuole-like and punctate structures in yeast, reticulated cytoplasmic texture and high-RI puncta in HepG2, and intracellular puncta in K562.

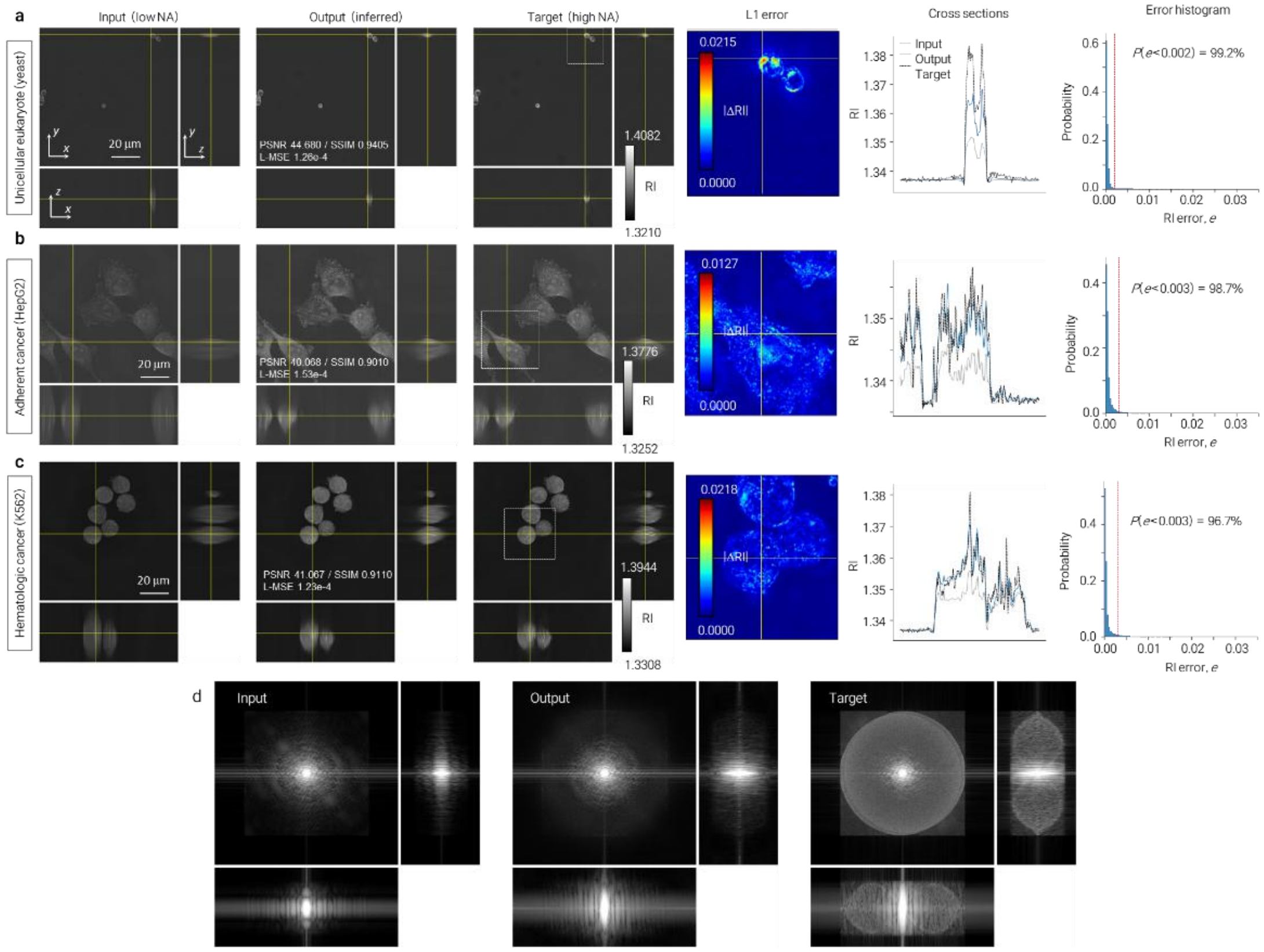

**Fig. 3. Cross-cell-type high-NA inference and error analysis.** ResShift-ODE was evaluated on three representative cell morphologies: (a) yeast, (b) HepG2, and (c) K562. For each class, orthogonal RI sections compare the low-NA input, inferred high-NA-equivalent

output, and experimentally measured high-NA target. Yellow and blue lines mark the positions used for cross-section and line-profile analysis, and reconstruction metrics are shown in the output panels. Enlarged regions show voxel-wise L1 error maps, $|\hat{x}_H - x_H|$, RI line profiles, and voxel-wise error histograms. (d) Fourier-domain amplitude projections show recovery of spatial-frequency content beyond the low-NA support without artificial filling of the axial missing cone.

The inferred outputs recover much of this high-NA-like morphology without per-class fine-tuning. A single model transferred across the three morphologies tested: suspended, adherent, and organelle-rich suspension, which is consistent with, but does not by itself demonstrate, a prior anchored to the shared coherent forward operator rather than to cell-type-specific image statistics. Because all three classes were also represented in training, genuine transfer to held-out cell types remains untested and is left to future leave-one-class-out evaluation. Across all three held-out volumes, cell boundaries sharpen, internal RI texture becomes more continuous, and the *xy*/*yz*/*xz* views remain mutually consistent. These assignments are morphological and RI-based, not label-confirmed, so organelle-level labels in the inferred output should be interpreted cautiously.

Voxel-level comparisons support the visual result. The error histograms in Fig. 3 show that 99.92% of yeast voxels fall within RI error 0.002, while 99.39% of HepG2 voxels and 99.51% of K562 voxels fall within RI error 0.003. Lateral RI profiles likewise overlay the high-NA targets across cell bodies and subcellular peaks. Because these values are computed on emulated low-NA inputs, they quantify inversion fidelity for the analytic operator, not independent validation on physical low-NA acquisitions. They are also dominated by the preserved low-NA support, where recovery reduces to passing through measured frequencies; the decisive quantity is recovery accuracy in the annulus outside the low-NA support, which we examine spectrally in Section 3.3.

Residual error is concentrated at thin, high-RI-gradient interfaces, most visibly around nuclear-envelope-like transitions in HepG2. This localization is consistent with the physics of the forward operator: sharp interfaces carry the largest fraction of their energy in the high-frequency band that the low-NA operator annihilates, so they are precisely the structures least constrained by the measurement and most dependent on the prior. The model under-recovering rather than over-sharpening these edges is therefore the expected signature of measurement-anchored completion reaching the edge of what the low-NA data constrain (Section 3.3), itself evidence against indiscriminate synthesis, rather than a global failure of cell-scale recovery, and it defines the principal structural limitation of the current model.

### *3.3 Fourier-domain consistency and measurement anchoring*

The Fourier-domain panels in Fig. 3 provide the key physical check on hallucination risk. The model output expands the lateral high-frequency support beyond the low-NA input but remains below the high-NA target, rather than filling all missing frequencies indiscriminately.

Importantly, the output does not artificially fill the axial missing-cone region. We distinguish two Fourier regions: (i) the lateral high-frequency annulus outside $\mathcal{K}_L$ but within $\mathcal{K}_H$, which is populated in the high-NA targets but absent from the low-NA measurements and is the intended target of NA transfer; and (ii) the axial missing cone, which arises from the diffraction-tomography geometry and which the high-NA reference $x_H$ itself does not populate, since it is common to both NA configurations [19]. Spectral power in the axial cone remains close to the low-NA input level for input, output, and target alike, whereas the lateral annulus is populated in a measurement-anchored manner. This asymmetry has a concrete origin in the training construction: because every pair is produced by the same operator that annihilates the axial cone, and the residual-shifting trajectory is anchored to the low-NA volume as its endpoint, the prior is never presented with an example in which cone energy must be synthesized from a low-NA input. It therefore learns to complete the lateral annulus populated in the high-NA targets while leaving the axial-cone region near the low-NA input level. The output should thus be read as a learned plausible completion of operator-annihilated lateral frequencies, constrained by the measured low-NA support, not as a direct measurement of the unobserved band.

### *3.4 Inference cost and baseline comparison*

We next compared the inference cost and reconstruction metrics of ResShift-ODE with three baselines: a 3D U-Net, a 1000-step 3D denoising diffusion probabilistic model (DDPM), and stochastic 3D ResShift (Table 1). The four-model comparison was performed on the HepG2 held-out volume. To show that the proposed model behaves consistently across the other held-out specimens, the lower rows report ResShift-ODE results on the yeast and K562 volumes. Wall-clock time, relative cost, and number of function evaluations (NFE) are measured per volume and represent model-level inference cost, whereas PSNR, SSIM, and L-MSE are single-volume reconstruction metrics.

**Table 1. Inference cost and reconstruction metrics on held-out test volumes.**

| Model | Relative wall-clock cost | Wall-clock | Steps (NFE) | PSNR (dB) | SSIM | L-MSE |
|---|---|---|---|---|---|---|
| 3D U-Net | 1× | 37.86 s | 1 | 39.677 | 0.9030 | 1.50e-4 |
| 3D DDPM | ~954× | 36128.62 s | 1000 | 38.875 | 0.8853 | 4.89e-4 |
| 3D ResShift | ~15× | 579.01 s | 15 | 39.923 | 0.8918 | 1.51e-4 |
| **3D ResShift-ODE (ours)** | **~5.8×** | **217.98 s** | **5** | **40.068** | **0.9010** | **1.53e-4** |
| ours, Yeast | ~5.8× | 217.98 s | 5 | 44.680 | 0.9405 | 1.26e-4 |
| ours, K562 (lymphoblast) | ~5.8× | 217.98 s | 5 | 41.067 | 0.9110 | 1.23e-4 |

ResShift-ODE required five denoiser evaluations and 217.98 s per 1092 × 1092 × 36 RI volume on an NVIDIA A100-SXM4-80GB GPU. This corresponds to ~5.8× the wall-clock cost of the deterministic 3D U-Net, but it reduces inference time by ~166-fold relative to the 1000-step 3D DDPM baseline. Compared with stochastic 3D ResShift, which requires 15 denoiser evaluations, ResShift-ODE reduces the number of evaluations to five and gives a ~2.7-fold wall-clock speedup by removing the Brownian sampling term. Thus, the main computational advantage of ResShift-ODE is deterministic few-step inference with substantially reduced sampling cost compared with stochastic diffusion baselines.

On the HepG2 held-out volume, ResShift-ODE achieved PSNR, SSIM, and L-MSE values comparable to the deterministic 3D U-Net and stochastic ResShift. Because these scalar metrics are computed on a single held-out volume and are dominated in part by the preserved low-NA passband, small differences below approximately 1 dB in PSNR or 0.01 in SSIM should not be interpreted as a statistical ranking of reconstruction accuracy. Instead, the metrics indicate that ResShift-ODE maintains full-volume reconstruction fidelity while providing deterministic diffusion-prior inference at a substantially lower cost than stochastic diffusion sampling.

These observations support the use of ResShift-ODE not as a scalar-metric replacement for a deterministic regressor, but as a measurement-anchored generative prior for completing the frequency band absent from the low-NA measurement.

## 4. Discussion

The central result of this study is that a residual-shifted diffusion prior, reformulated as a deterministic ODE sampler, can invert an analytically defined low-NA HT operator while retaining physically interpretable Fourier behavior. On held-out emulated low-NA/high-NA pairs, the model recovers high-NA-like cell boundaries and internal RI structure across yeast, HepG2, and K562 morphologies. The voxel-level results quantify inversion of the specified low-NA forward operator; they are not population-level statistics and do not yet establish quantitative performance on independently acquired physical low-NA data. The key assumption is that the numerical low-NA emulation captures the dominant information loss of a real low-NA acquisition. If it fails, the method still demonstrates operator inversion under controlled physics, but practical low-NA deployment would require a registered real-low-NA/high-NA validation set.

A residual-shifted diffusion prior is a natural fit for this inverse problem. A deterministic 3D U-Net can regress efficiently but tends to smooth high-frequency structure, whereas stochastic or weakly conditioned generative models can synthesize plausible but measurement-inconsistent details [36,37,45]. ResShift anchors the forward process to the low-NA endpoint rather than to an unconditional Gaussian prior [34], and the ODE reformulation removes Brownian sample-to-sample drift, making the reconstruction reproducible and conditioning it on the observed low-NA volume.

Recent work has shown that learned priors can substantially improve computational microscopy and RI tomography, but the role of the prior depends on the physical information gap being addressed. In 2D microscopy, diffusion models have been used with incoherent PSF models for image reconstruction [26] and with Brownian-bridge formulations for virtual staining [27]. In 3D RI tomography, learned or deterministic priors have been coupled to reconstruction algorithms to reduce missing-cone or limited-angle artifacts [21] [22], and slice-wise diffusion priors have been used for general 3D inverse problems with per-volume guidance [33]. These studies highlight the usefulness of learned priors for underdetermined optical inverse problems. In this sense, the contribution is not a new diffusion sampler alone, but a physics-defined use of a deterministic residual-shifting prior for NA transfer: the prior is asked to complete only the lateral band missing from low-NA acquisition while preserving the Fourier regions that should remain unfilled.

The main safeguard against unconstrained synthesis is measurement anchoring. Because the reverse trajectory is conditioned on the low-NA measurement and the output is evaluated in Fourier space, recovered structures can be interpreted according to where their spectral content lies. Features within the shared low-NA support remain directly constrained by the measurement, whereas features in the recovered lateral high-frequency band should be treated as prior-guided estimates. This distinction is important for downstream biological analysis, where quantitative readouts should separate measured RI information from inferred high-frequency content.

The computational cost is also compatible with post-acquisition screening workflows. ResShift-ODE requires five denoiser evaluations and ~218 s per 1092 × 1092 × 36 volume, which is substantially faster than the 1000-step DDPM baseline and faster than stochastic ResShift. At multiwell-screening scale, this deterministic few-step inference is more practical than long stochastic sampling chains, which would require ~10 h per volume in our implementation. The acceleration arises from the shortened residual-shifting trajectory of ResShift [34] and the removal of the Brownian sampling term by the ODE sampler. Further speedups through distillation [47,48], quantization, or optimized ODE schedules should make this approach more suitable for plate-scale deployment.

This principle is not limited to scalar RI recovery in static HT. HT variants with a defined acquisition operator and reproducible information gap could also benefit from measurement-anchored prior completion, including spectroscopic HT [49], dielectric tensor tomography [50,51], and flow-cytometry HT [52]. Each extension would require modality-specific forward operators and calibration data, but the core principle—prior-guided recovery constrained by measured support—remains the same.

Several limitations remain and define the next validation steps. The present quantitative results rely on numerically emulated low-NA inputs, so registered physical low-NA/high-NA data, or registration-invariant Fourier-domain metrics, will be needed to establish real-world performance. Residual errors are concentrated at thin high-RI-gradient interfaces, suggesting that future training should emphasize high-frequency boundary recovery through frequency-aware augmentation, edge-weighted losses, or membrane-enriched examples. The current dataset is also limited to three cell types acquired under air-immersion HT on a single instrument family; extension to organoids, tissue slices, strongly scattering samples, and water- or oil-immersion regimes will require modality-specific forward models and paired calibration data. Beyond validation, the deterministic ODE trajectory could support uncertainty estimation, confidence-guided segmentation, and joint learning of NA transfer with aberration correction.

## 5. Conclusion

We have presented a 3D generative framework for holotomography that transfers low-NA (0.38) to high-NA (0.72) RI volumes by deterministic inference under a ResShift residual-shifting diffusion prior, reformulated as a probability-flow ODE. On held-out emulated-pair test volumes, the model recovers high-NA-like 3D RI structure across yeast, HepG2, and lymphoblasts (K562), with more than 99% of voxels within RI error 0.002–0.003 per class (one held-out volume each; Section 3.2), a figure dominated by the preserved low-NA passband and therefore a measure of operator-inversion fidelity, not of recovered high-frequency content. Fourier analysis shows lateral high-frequency recovery without filling the axial missing cone, and the five-step ODE sampler reduces inference cost relative to stochastic baselines.

More broadly, the construction, a deterministic residual-shifting ODE prior anchored to a measurement under an explicit, known band-limiting operator, offers a template for coherent inverse problems in which a defined Fourier region is annihilated rather than attenuated, such as limited-angle tomography, synthetic-aperture extension, and sub-pupil coherent imaging, with preservation of the unmeasured missing cone serving as a falsifiable, modality-agnostic acceptance criterion. These findings position ResShift-ODE as a software route toward improving the resolution-throughput trade space of existing holotomography systems without changing the acquisition hardware. The current quantitative evidence is strongest for inversion of the analytically defined low-NA operator; physically acquired low-NA performance remains to be established. With registered real-low-NA/high-NA validation, the approach could let multiwell- and organoid-compatible screening optics recover laterally resolved RI features: subcellular puncta-like structures, internal texture, and sharpened boundaries (morphological and RI-based rather than label-confirmed; Section 3.2), features that label-free morphological profiling and cell-state scoring depend on but that today require short-working-distance high-NA optics—relevant to high-content drug-response phenotyping and cell-therapy quality control of unstained live products. Realizing this requires registered physical low-NA/high-NA validation across multiple specimens per class and quantitative out-of-band recovery metrics, neither of which the present emulated, single-volume-per-class evidence provides.

**Funding.** Research Foundation of Korea (NRF) grant funded by the Korean government (MSIT) (RS-2024-00442348, RS-2022-NR068141, RS-2024-00448562, RS-2024-00451981, RS-2024-00440577); Korea Institute for Advancement of Technology (KIAT) through the International Cooperative R&D Program (P0028463); Korean Fund for Regenerative Medicine (KFRM) grant funded by the Ministry of Science and ICT and the Ministry of Health & Welfare RS-2024-00332454); the Korean Fund for Regenerative Medicine (KFRM) grant funded by the Korea government (the Ministry of Science and ICT and the Ministry of Health & Welfare) (21A0101L1-12).

**Acknowledgment.** We thank Prof. Sung-Bin Lim (Korea University) for early-stage methodological discussions on accelerated diffusion sampling, on incorporating cell-prior distributions, and on Fourier-space-aware sampling such as cold diffusion, which shaped the design space for the ODE reformulation reported here. Tae-Young Park (UNIST) contributed an early 3D DDPM exploration, reported separately [53].

**Author contributions.** H.-S.M. and Y.K.P. conceived the project. D.-M.R., J.-H.R., and H.-S.M. designed the methodology. D.-M.R. and J.-H.R. developed the software, performed validation and visualization, and curated the data together with W.-S.L. and H.Y., who also performed the holotomography acquisitions. Y.K.P., D.-M.R., and J.-H.R. drafted the manuscript; all authors contributed to review and editing. H.-S.M. and Y.K.P. supervised the project and led project administration; Y.K.P. acquired funding.

**Disclosures.** Y. K. P. is a co-founder and shareholder of Tomocube Inc., the manufacturer of the HT-X1 system used in this study. D.-M. R., J.-H. R., W.-S. L., H. Y., and H.-S. M. are employees of Tomocube Inc. The authors declare no other competing interests.

**Data availability.** The paired Low-NA/High-NA RI volumes that support the findings are available from the corresponding authors upon reasonable request.